\documentclass[aps,prb,twocolumn,superscriptaddress]{revtex4-1}
\usepackage{graphicx}
\usepackage{amssymb}
\usepackage{amsthm}
\usepackage{amsmath}
\usepackage{rotating}
\usepackage{color}
\usepackage[colorlinks,citecolor=blue,linkcolor=blue]{hyperref}
\def\bl#1{\textcolor{black}{#1}}

\begin{document}
\title{Grain Boundary Premelting of Monolayer Ices in 2D Nano-channels}
\author{Zun Liang}
\affiliation{Physics Department, School of Physics and Material Science, East China Normal University, Shanghai 200062, China}
\author{Han Du}
\affiliation{Physics Department, School of Physics and Material Science, East China Normal University, Shanghai 200062, China}
\author{Hongtao Liang}
\thanks{52150602013@stu.ecnu.edu.cn}
\affiliation{Physics Department, School of Physics and Material Science, East China Normal University, Shanghai 200062, China}
\author{Yang Yang}
\thanks{yyang@phy.ecnu.edu.cn}
\affiliation{Physics Department, School of Physics and Material Science, East China Normal University, Shanghai 200062, China}
\date{\today}
\begin{abstract}
We employ molecular-dynamics (MD) simulations to study grain boundary (GB) premelting in ices confined in two-dimensional hydrophobic nano-channels. Premelting transitions are observed in symmetric tilt GBs in monolayer ices and involve the formation of a premelting band of liquid phase water with a width that grows logarithmically as the melting temperature is approached from below, consistent with the existing theory of GB premelting. \bl{The premelted GB is found rough for a broad range of temperature below the melting temperature, the two ice-premelt interfaces bounding the melted GB are engaged with long wave-length parallel coupled fluctuations.} Based on current MD simulation study, one may expect GB premelting transitions exist over a wide range of low dimensional phases of confined ice and shows important consequences for crystal growth of low dimensional ices.
\end{abstract}
\maketitle
The term premelting refers to the formation of a thermodynamically stable liquid film at solid interfaces at temperatures below but close to the bulk melting temperature $T_m$.~\cite{Chernov93} The premelting transitions occur in all types of solids, they stand out in the case of ice, because of their association with terrestrial life and the importance of their environmental effects.~\cite{Dash06} Studies of premelting at ice surfaces have been carried out with a wide variety of modern experimental and molecular-level modeling techniques (see in Ref.[\onlinecite{Dash06,Bjorneholm16}] and references therein), these studies have demonstrated the rich nature and complexity of the ice surface premelting behaviors, which covers a span from quasi-liquid bilayer-by-bilayer melting process~\cite{Sanchez17,Pickering18} to complete,~\cite{Elbaum93} incomplete premelting~\cite{Elbaum91} and a first-order phase transition between two distinct premelting states,~\cite{Murata16} depending on temperature, surface crystallographic orientation and even atmospheric environment.

In comparison to surface premelting, the complexity mentioned above is only the beginning to be explored for grain boundary (GB) premelting of ices, as more bicrystallography parameters (symmetry, misorientation, and boundary plane) are introduced. It is well known that GB premelting of ice may play a vital role in a variety of processes of geophysical, geological and atmospheric interest.~\cite{Dash06} However, the challenges inherent in characterizing the structure of ``buried'' internal GBs have significantly limited the number of direct experimental studies,~\cite{Thomson13} and have resulted in a situation that GB premelting is one of most important yet the least studied aspects of ice. During the last decade, GB premelting of pure metals and binary alloys, have been the subject of continuum modeling studies~\cite{Tang06a,Tang06b,Berry08,Mellenthin08,Mishin09b,Cogswell11} and atomistic simulations studies,~\cite{Hoyt09,Williams09,Frolov09,Fensin10,Mishin10,Olmsted11,Hickman16} which led to insights into the nature of the thermodynamic driving forces of the GB premelting as a function of GB bicrystallography. In the meantime, advances in the modeling and simulation of ice GB premelting remain scarce.~\cite{Prinzio16,Moreira18}

The present study is also motivated by a recent high-resolution electron microscopy study by Algara-Siller et al.~\cite{Algara-Siller15} In their study, the two dimensional (2D) GBs between the crystallites of monolayer ice which adopts unexpected simple square lattice, inside the hydrophobic bi-graphene nano-capillaries, were directly observed. Although no sign of GB melting was found within specific setup in the experiments, their observations still provide a compelling brand-new perspective for the exploration of GB premelting of ices. 

In this paper, we report molecular-dynamics (MD) simulations on a simple symmetric tilt GB of monolayer ices confined in 2D hydrophobic nano-channels that predict the existence of premelting transition at such low dimensional ice GB. In this process, a premelting liquid layer (or ``band'' due to its 2D nature) forms at the GB between two confined monolayer ice crystallites below $T_m$ of their 2D ``bulk'' phase. To the best of our knowledge, GB premelting has not been previously reported within the 2D or semi-2D confined ices, either by experimentally or by simulation. As we will demonstrate, the width of the premelting band depends logarithmically on the undercooling, indicating the process of this specific 2D GB premelting system is repulsive. \bl{Besides, the parallel capillary waves of the two coupled crystal-remelt interface bounding the melted GB are explored, and we identify the rough nature of current low dimensional premelted GB, inconsistent with bulk ice surface premelting which needs to reach the high homologous temperature to undergo roughening transition.} As a consequence, it is expected that this lower dimensional GB premelting phenomena could potentially serve as a testing ground to gain molecular-level insights for understanding the complicated GB premelting of 3D bulk ices.

GB premelting occurs near $T_m$ when the interfacial free energy of a dry GB, $\gamma_{GB}$, is larger than twice the solid-liquid (solid-premelt) interfacial free energy (2$\gamma_{SL}$), $\Delta \gamma =  \gamma_{GB} - 2\gamma_{SL} > 0 $. When the undercooling ($\Delta T = T_m - T$) is not too great, the increase in bulk free energy is greater than compensated by the decrease in the total interfacial free energy, a thin metastable quasi-liquid layer is thus thermodynamically favorable. The variation of the width of the quasi-liquid layer $w$ depends on $\Delta \gamma$ and undercooling $\Delta T$, and the short range structural force (referred as ``disjoining potential'') between the two solid-premelt interfaces bounding the premelted layer. In 1980, Kikuchi and Cahn~\cite{Kikuchi80} predicted a logarithmic dependence of the variation of $w$, as a function of $\Delta T$ for GB premelting,
\begin{equation}
w(\Delta T)= -w_0 \ln{[\Delta T/T_0]},
\label{eq:wlog}
\end{equation}
$T_0$, $w_0$ are constants specific to the given GB. In the derivation of Eq.~\ref{eq:wlog}, an assumption that the disjoining potential is exponential and repulsive was employed, $g(w) = \Delta \gamma e^{-w/w_0}$, where $w_0$ is the decay length of the disjoining potential $g(w)$. $T_0 = \Delta \gamma T_m/w_0 \rho L$, $\rho$ is the particle number density of the premelted liquid, $L$ is the latent heat.

To model confined monolayer ice, we follow part of the simulation setup by Algara-Siller et al. in which a hydrophobic nano-channel consist of two walls under a fixed separation of 6.5\AA was employed to accommodate monolayer ice.~\cite{Algara-Siller15} In their study, simulation results for monolayer ice showed little dependence on water models, i.e., SPC/E and TIP4P/2005, so we choose to employ only SPC/E in our simulations. Besides, to mimic the squeezing effect on the encapsulated water molecules due to the adhesion (van der Waals) force between the encapsulating walls, we apply a lateral pressure of 0.5GPa as Algara-Siller et al. did to the confined ice. The water-wall interaction is modeled using a simple pairwise potential,\cite{Magda85} in which the parameters were chosen as in Ref.~\onlinecite{Du18}. Under these confinement parameters, we have previously reported (from MD) the phase transition between the monolayer ice and corresponding liquid phase is first-order, and a precise melting point ($T_m$=402.5K) determined from a delicate crystal-melt phase coexistence simulation.~\cite{Du18} \bl{Note that $T_m$ for bulk ice modeled with SPC/E was reported as 215K,~\cite{Abascal07,Vega11} nearly 200K smaller than that of the monolayer square ice in the current study, such big difference in $T_m$ between the 2D and 3D ice is not surprising. It has been known that the phase behavior (such as melting point) of the square-like confined ice is highly depending on confinement width as well as the lateral pressure.~\cite{Algara-Siller15,Chen16,Zhu18}}

The molecular dynamics simulations are performed using open source software LAMMPS released by the Sandia National Lab.~\cite{Plimpton95} The long-range interactions are computed using the particle-particle-particle-mesh (PPPM) algorithm. Periodic boundary conditions are applied in $x$ and $y$, while in $z$ empty volume between simulation boxes are inserted to remove the inter-box interactions.~\cite{Yeh99} By tilting two periodic square ice lattice and merging them, an initial configuration containing a pair of symmetric tilt GBs with misorientation angle $\theta=30^\mathrm{o}$ is created to perfectly fit the $xy$ periodic cell, as illustrated in the schematic diagram in Fig. 1. We use $xy$ cell dimension ($L_x\times L_y$) of approximately 750\AA $\times$46\AA, containing around 13,000 water molecules.

\begin{figure} [!htb]
\includegraphics[width= 3.5in]{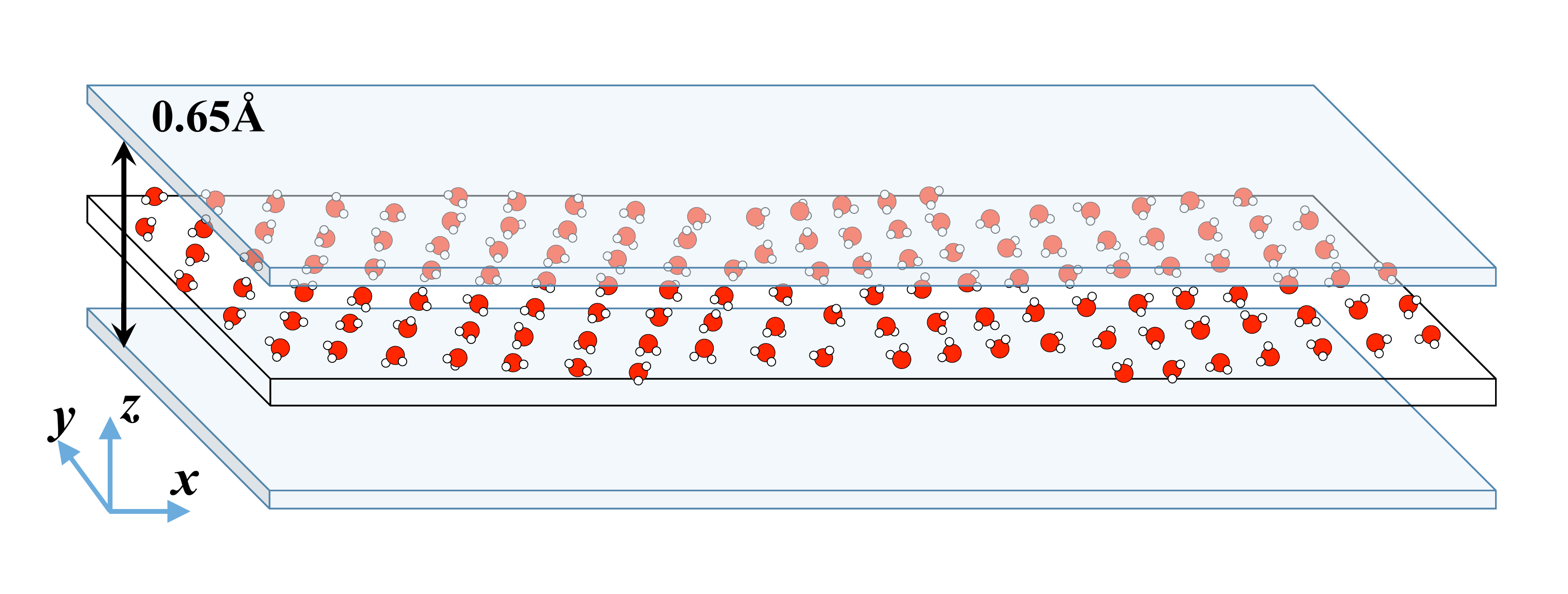}
\caption{Schematic diagram of the simulation cell. A pair of symmetric tilt GBs of the monolayer ices are confined inside a hydrophobic nano-channel consist of two walls under a fixed separation of 6.5\AA.}
\label{fig1}
\end{figure}

Equilibrated GB interfaces are set up at varying temperatures $T$ ranging from 360 to 400 K, separated by 10 K. To produce the equilibrated GBs structure, the constant area and lateral pressure MD ($NP_xAT$) simulations up to 10 ns are employed to yield equilibrium number density and pressure (0.5GPa). These setup are followed by 5ns $NVT$ simulations to collect data. Time step of 1.0 fs is used. Eight replica systems (each containing two independent GBs) are used at each $T$ to improve statistics. Several profiles that show the variation of temperature and pressure (including three tensor components) along $x$-axis, were examined for all simulations to confirm the absence of temperature gradients and residual stress in ``bulk'' square ice.

The equilibrium GBs are characterized through the determination of density map which shows the change in areal density as functions of the $x$ and $y$ direction. We also calculate the order-parameter (OP) profiles, which uses a local structure OP that distinguishes 2D ice from water phase,~\cite{Morris02} to determine the extent of the premelting region,  for each single MD trajectory. Then the instantaneous GB width, $w_t$ is defined as the half value distance at the profile, see in Fig. 2 (b), and the equilibrium premelting width $w$ is calculated by averaging all $w_t$ over time. \bl{By choosing a threshold OP value of 0.32 to distinguish liquid like from solid like molecules (in the range of temperatures studied) and by defining coarse-grained grids the $xy$ plane (with grid size $\sim$3\AA$\times$3\AA), the positions of the two interfaces between solids and the premelt film could be located. The positions of the two interfaces, $h_{sp}(y)$ for the solid-premelt interface and $h_{ps}(y)$ for the premelt-solid interface, are determined as the average of the $x, y$ coordination of all the outermost solid-like molecules contained within each corresponding bin. Knowledge of the time evolution of the $h_{sp}(y)$ and $h_{ps}(y)$ can be used to calculate the spectrum of coupling fluctuations directly as well as to provide an assessment of the validity of the coupled capillary wave theory theory~\cite{Benet16} in the 2D or semi-2D confined ice/premelt system.}

\begin{figure} [!htb]
\includegraphics[width= 3.5in]{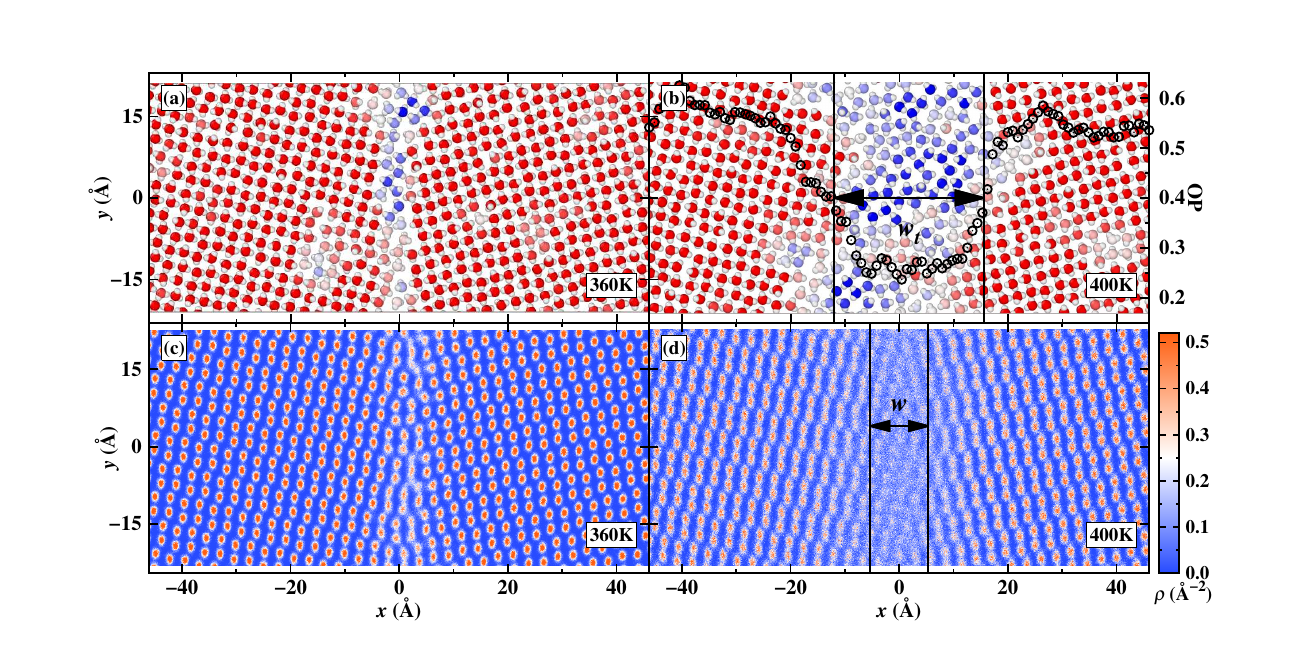}
\caption{Snapshot of a single symmetric tilt GB of confined monolayer ice, misorientation angle $\theta=30^\mathrm{o}$. (a) $T=$360K, GB becomes increasingly disordered while the ordered bridges can still be recognized. (b) $T=$400K, 2.5K below $T_m$, GB premelts. Oxygen atoms are colored coded on the structural OP as ice (red) and premelted liquid (blue). The corresponding order parameter profile used to calculate the instantaneous GB width $w_t$ is also plotted on top of the snapshot. (c) and (d), color-contour plots of the time-averaged areal density map for the premelting GBs of the confined monolayer ices under 360K and 400K, respectively. Red color stands for density peaks while blue for zero value of areal density. A double arrow in (d) represents the magnitude of the equilibrium premelting width $w=10.7\AA$.􏰠}
\label{fig2}
\end{figure}

\begin{figure} [!htb]
\includegraphics[width= 3.5in]{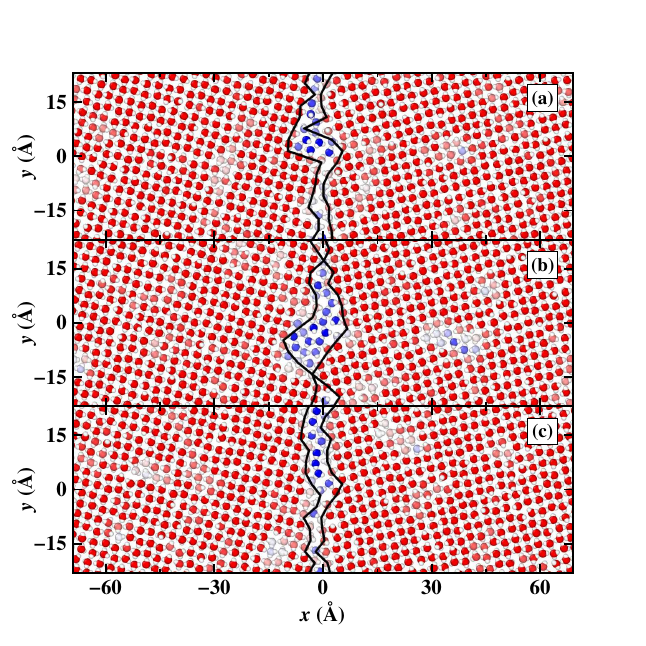}
\caption{\bl{(a), (b) and (c) show successive trajectories of premelted GB with a 10ps time interval at $T=$360K. Oxygen atoms are colored coded on the structural OP as ice (red) and premelted liquid (blue). The calculated two solid-premelt interfacial position profiles along $y$ axis are plotted as black solid lines.}}
\label{fig3}
\end{figure}

\bl{Fig. 2 (a) shows a $NVT$ snapshot of an equilibrated monolayer ice GB at $T=$360K. The oxygen atoms have been color-coded based on their OP values, blue representing a liquid-like environment and red the monolayer ice crystal. Panel (a)-(c) in Fig. 3 show a series of snapshots (separated by 10 ps) of the same GB in Fig. 2 (a). At this temperature, the structural disorder in the grain boundary is nonuniform and dynamic. Certain parts of the grain boundary become tiny pockets with liquid-like water molecules rotate and swap positions, while ordered solid bridges forms in the remaining portions of the boundary, see in Fig. 3(b). However, we find these order-disordered structures are highly dynamic, solid bridges form and disappear rapidly on the time scale of the simulations. Such behavior is rather consistent with premelting behavior at 3D symmetric tilt GB in metal Ni under large undercooling.~\cite{Fensin10} The two solid-premelt interfacial positions $h_{sp}(y)$ and $h_{ps}(y)$, determined from discretized bins, are represented by two thick solid lines, which could well depict the liquid pocket and the solid bridge just mentioned. The rugged shape of the solid-premelt interfacial position indicates its rough nature in molecular level, noticeable instantaneous kinks yet no evident overhangs are found along the interfaces. By tracing the time evolution, we find the two interfaces fluctuate cooperatively and are highly correlated, and sometimes the whole GB (including the two parallel coupled solid-premelt interfaces) could become curved (as shown in Fig. 3(a)), despite of the fact that current 2D simulation system only has a dimension of $L_y=46\AA$. Overall, the width of the liquid-like premelt is very small at 360K. Usually, only one or two liquid-like water molecule exist between the two interfaces, see in Fig. 3(c).}

\bl{At intermediate temperatures, $T=370K\sim390K$, the premelted grain boundary turns into a 2D band, in which the water molecules possess highly disordered structures and actively swap positions with neighbors. As the temperature rises, e.g., from Fig. 4(a) to (c), the liquid-like disorder increase gradually and propagate outward, the width of the liquid band is around two diameters of the water molecules at 370K, and reaching to a value of more than three diameters at 390K. The vacancy defect sometimes forms at the solid-liquid interfaces and diffuses into the crystals as the temperature is approaching $T_m$. Instantaneous kinks at square ice surfaces become less but could still be found, the increase in the width of the liquid-like band weakens the coupling of the two interfaces. However, the two interface fluctuations are still positively correlated. By examining the successive time evolution of the premelting GB, we find a distinct mode of mass transport. Small clusters of water molecules collectively lose their crystalline order and collectively transform to the crystalline phase, either back to their original grain or attach to the opposite grain. Such frequent inter-detachment-attachment processes are associated with inter-diffusion and inter-mixing between point vacancy defects and water molecules at the two interfaces.}

\begin{figure} [!htb]
\includegraphics[width= 3.5in]{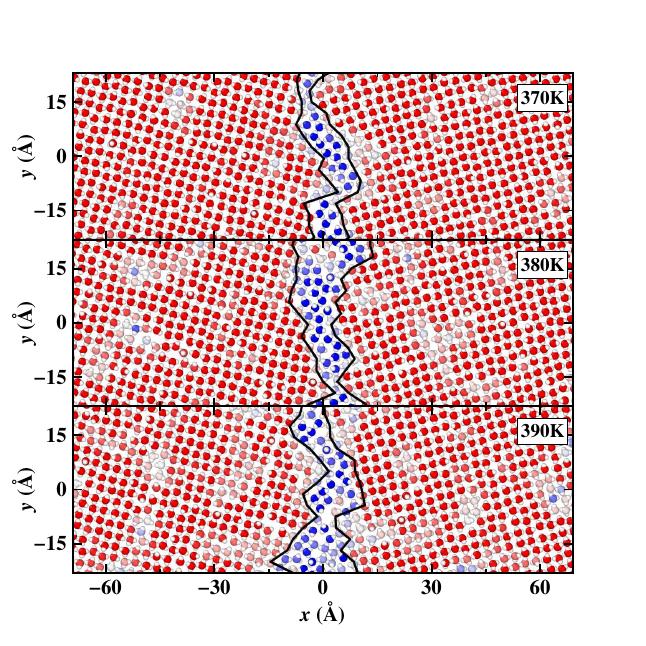}
\caption{\bl{Snapshot of premelted symmetric tilt GBs of confined monolayer ice, misorientation angle $\theta=30^\mathrm{o}$, at intermediate temperatures, $T=$370K (a), $T=$380K (b) and $T=$390K (c). Two calculated solid-premelt interfaces bounding premelting region are plotted as black solid lines.}}
\label{fig4}
\end{figure}

\bl{At temperature approaching the $T_m$, i.e. $T=$400K. The disorder and rapid diffusion are further increased, and the premelted band between the two grains of monolayer square ice grows even thicker. The premelted regions are highly dynamic in both instantaneous solid-liquid interfacial positions and the position of the center of mass of the liquid-like band. The panel of Fig.2 (b) demonstrates the presence of premelted liquid phase between two solid-premelt interfaces. Also plotted is the corresponding structural order parameter profile, with an instantaneous premelting width $w_t$ reaching nearly 30\AA. Because of the large separation between the two interfaces, the parallel correlation between the two interfaces decreases significantly, and the curvature of the whole GB at this temperature is hardly visualized due to the short system dimension employed in the current study.} 

Contour plots of the time average oxygen density map are shown in Fig. 2(c) and (d) for the GBs at 400K and 360K, respectively. The peaks of the density are seen as red in the plot, which are stronger and more highly localized in both ice grains. For a relative dry boundary at 360K, density  exhibits oscillatory behavior. Whereas densities of the premelted boundary at 400K is nearly free of density peaks which appears as white color in the contour plot, the individual dislocations cannot be distinguished. The premelting is found to proceed by the uniform homogeneous band along the boundary rather than dislocation-pairing transition which was previously found in symmetric tilt GBs for BCC Fe.~\cite{Olmsted11}

\bl{To further verify the liquid nature of the premelting region, we examine both the local structure and the transport of the water molecules within the premelting band. The local structure in premelt is examined by analyzing the two-dimensional (in-plane) oxygen-oxygen radial distribution function (RDF), $g_{2d}(r)$, only for the water molecules within a dynamic bin, $\pm1\AA$ away from the instantaneous middle of the premelting band, which is defined as half of $(h_{sp}+h_{ps})$. For comparison, we have also carried out a two-dimensional RDF analysis for the 2D bulk liquid monolayer water under the same temperature. Fig. 5 (a) and (b) show the calculated two-dimensional O-O RDF for the premelts at 390K and 400K, respectively. For 390K, only the first $g_{2d}(r)$ peaks show a minor difference in amplitude compared with the respective 2D bulk liquid phase. Since the average premelting width for 390K no more than 6\AA, significant solid-like structure start to show up when $r>4\AA$. Fig. 5 (b) plots $g_{2d}(r)$ results at 400K, peaks $g_{2d}(r)$ are commensurate with those of their respective bulk liquid phases better than 390K. Results of two different premelting replica MD runs are shown, for the one replica with thicker $w$ ($\sim16\AA$), $g_{2d}(r)$ is entirely consistent with the structure of bulk 2D liquid phase, as illustrated with the thicker dashed line in Fig. 5(b). The transport is characterized through the calculation of two-dimensional (in-plane) mean-squared displacement (MSD) of the same amount of water molecules in the middle of the premelting band versus time.\cite{Yang13} The MSD used in this analysis are calculated out to 20 ps, which is sufficient time for the dynamics to become diffusive, but smaller than the average time required for a molecule to diffuse halfway across the premelting band, we also abandon those particles freeze within each 20 ps based on the structural order parameter. The dashed line in Fig. 5(c) and (d) represents the MSD results for those selected water molecules at premelted GBs under $T$=390K and 400K. For 400K, the magnitude of slope of the diffusive regime of the MSD gives a in-plane diffusion constant $D_{xy}$=$2.45(11)\times 10^{-5}$cm$^2$s$^{-1}$ is identical to the monolayer bulk liquid value of $2.48(14)\times 10^{-5}$ cm$^2$s$^{-1}$, where values in parentheses represent 95\% confidence level error estimates in the last digits. For 390K, small difference in the slope of the diffusive regime in MSD comparing premelt and bulk phase, could due to the correlation between the liquid-like molecules in the premelt band and the nearby crystal surfaces. Overall, results in Fig. 5 suggest that the premelting band at 2D monolayer ice GB inherits more structural and transport property of monolayer bulk liquid, and shows less influence of either confinement or the nearby presence of crystalline environment, at higher $T$. The transport result is in contrast to premelting at GBs in 3D ice $\mathrm{I_h}$, in which transport of the water molecules within the premelting layer was predicted to behave sub-diffusively even at a temperature just below the $T_m$.~\cite{Moreira18}
}

\begin{figure} [!htb]
\includegraphics[width= 3.5in]{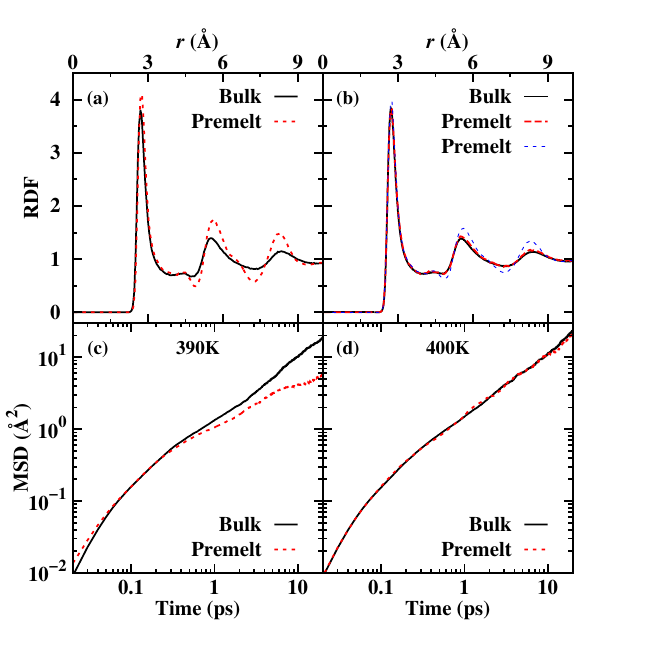}
\caption{\bl{(a) A comparison of the 2D O-O radial distribution function (RDF) between the undercooled monolayer bulk water (solid line) and premelted water (dashed line) at monolayer ice GBs, $T=390K$. (b) Same comparison as in panel (a), two dashed lines stand for RDF results from two replica simulation systems out of eight at $T=400K$. (c) Log-log scale mean-squared displacement (MSD) results for both undercooled monolayer bulk water and premelted water at monolayer ice GBs, $T=$390K. (d) Same comparison as in panel (c), $T=$400K.}}
\label{fig5}
\end{figure}

To examine the validity of Eq.~\ref{eq:wlog} for the 2D GB premelting transition in monolayer ice in the nano-channels, we plot the calculated time-averaged width of the premelting band, $w$, as a function of the undercooling, $\Delta T = T_m - T$ on a linear-log plot in Fig. 6. Eq.~\ref{eq:wlog} well describes the data for undercooling up to 42.5K - \bl{no deviations are seen when $w$ approaches atomic dimensions at low temperatures.} The divergence of the $w$ as the $T_m$ is approached is indicative of a complete GB premelting transition with repulsive disjoining potential. This repulsive nature could arises from short-ranged structural forces associated with the overlap of the density waves in the diffuse regions of the two bounding solid-liquid (solid-premelt) interfaces. Using a weighted least-squares linear regression over the temperature range from 360K to 400K, we obtain estimates for $w_0=2.17(9)$\AA, $T_0=166(12)$K, where values in parentheses represent 95\% confidence level error estimates in the last digits. We should note that for systems with dispersion forces, the logarithmic divergence of $w$ may not hold, and may change to either a finite threshold (so called ``incomplete'' premelting) or a power-law divergence.\cite{Dash06} However, the observed premelting widths in Fig. 6 do not exceed a \bl{number} of atomic diameters, short-range structural forces are dominant over long-range dispersion forces.

\begin{figure} [!htb]
\includegraphics[width= 3.5in]{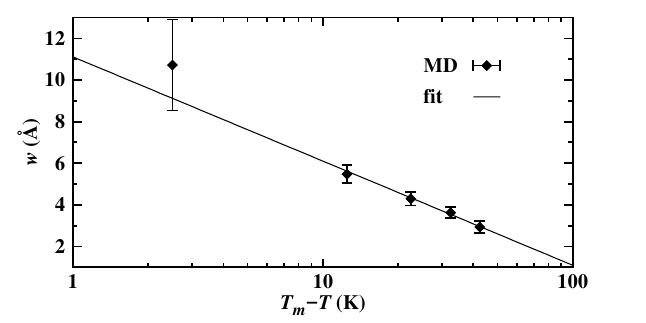}
\caption{Linear-log plot of the premelting width $w$ versus undercooling, $T_\mathrm{m}-T$. The solid line is the result of weighted least-squares fit to Eq.~\ref{eq:wlog}.}
\label{fig6}
\end{figure}

Hoyt and co-workers established a simulation-based methodology to accurately extract $g(w)$ from analyzing GB width fluctuations.\cite{Hoyt09} Their method had been applied to GBs in pure metals\cite{Fensin10} and binary alloys\cite{Hickman16} under various conditions (temperatures, misorientations, and chemical compositions). Due to the spatial oscillations of structural order and chemical compositional at two interfaces, $g(w)$ could display more complex dependencies on $w$ than suggested exponential decay form, i.e., attractive or intermediate potential with one or two local minima. Given the fact that there exist a great variety of low dimensional ice phases,~\cite{Algara-Siller15} we are optimistic that, the combination of the MD simulation setup (as presented here) with the analysis techniques described above can open numerous possibilities for investigating premelting low dimensional ices under various conditions, e.g., varying confinement length scale. Consequently, more complex shape disjoining potential can be anticipated in ice GBs, as found in metallic GBs. 

\bl{Benet et al.~\cite{Benet16} recently extended the variational theory for the sine-Gordon model by coupling capillary wave theory, and identified an important roughening transition in the premelted ice surface from low $T$ to high $T$. Their theoretical formulation leads to three relationships for the power-spectrum of interface fluctuations, in our case the three relationships reduce to two, and they are written as follows,}
\begin{equation}
\bl{
\langle |\hat{h}^2_{sp}(q_x)| \rangle=\langle |\hat{h}^2_{ps}(q_x)| \rangle=\frac{k_BT}{L_y} \frac{\omega + g^{''}+\tilde{\gamma} q_x^2}{\omega g^{''}+(2g^{''}\tilde{\gamma}+\omega\tilde{\gamma})q_x^2 +\tilde{\gamma}^2q_x^4}
}
\end{equation}
\begin{equation}
\bl{
\langle \hat{h}_{sp}(q_x)\hat{h}^*_{ps}(q_x)\rangle=\frac{k_BT}{L_y} \frac{g^{''}}{\omega g^{''}+(2g^{''}\tilde{\gamma}+\omega\tilde{\gamma})q_x^2 +\tilde{\gamma}^2q_x^4}
}
\end{equation}
\bl{
where $k_B$ is Boltzmann’s constant, $q_x = 2\pi n/L_x$, and $n$ is a positive integer, $\hat{h}_{sp}$ and $\hat{h}_{ps}$ are Fourier amplitudes of solid-premelt and premelt-solid interface fluctuations with wave vector $q_x$, respectively. $g^{''}$ is the second derivative of the disjoining potential $g(w)$ with respect to the premelting band width $w$, and $\tilde{\gamma}$ is macroscopic stiffness of the 2D ice-water solid-liquid interfaces in the limit of $q_x=0$. $\omega$ is a parameter that needs to be solved self-consistently and closely relates to the structural type of the whole interface/boundary. If $\omega\neq0$, the fluctuation amplitude remain finite at long wavelength limit, and the effective stiffness function $\Gamma_{\alpha\beta}(q_x)$ diverges as $q^{-2}_x$, the premelted GB is a smooth type boundary. Here the effective stiffness function is defined as,}
\begin{equation}
\bl{
\Gamma_{\alpha\beta}(q_x)=\frac{k_BT}{L_y\langle \hat{h}_{\alpha\beta}(q_x)\hat{h}^*_{\alpha\beta}(q_x)\rangle q_x^2}.
}
\end{equation}
\bl{However, if $\omega=0$, the two interfaces bounding the premelting band both behave as rough interfaces with the same effective stiffness, $\Gamma_{sp}(0)=\Gamma_{ps}(0)=\Gamma_{spps}(0)=2\tilde{\gamma}$, the whole premelted GB is rough type boundary.}

\bl{To test the validity of the Eq. (2)-(3) in premelted 2D square ice GB system, and to examine the roughness of both the GB and the two bounding solid-premelt interfaces. The three time-averaged power spectra of the capillary fluctuations are calculated with the knowledge of the time evolution of the solid-premelt interfacial position functions, $h_{sp}(y)$ and $h_{ps}(y)$. The results of effective stiffness functions obtained for the solid-premelt interface fluctuation spectra, for four temperatures, are plotted in Fig. 7. The up-pointing and down-pointing triangle symbols in the plots stand for the $\Gamma_{sp}(q_x)$ and $\Gamma_{ps}(q_x)$, two solid-premelt interfaces share the nearly identical shape in the fluctuation spectra. From 370K to 400K, the overall stiffness of the two bounding interface decreases as $T$ increases. For each temperature studied, over a range in $q$, $0.2<q<0.8$, constant values of $\Gamma_{sp}$ or $\Gamma_{ps}$ are found, this finding suggests that: a) fluctuation amplitudes within this $q$ range are perfectly scaled with $q^{-2}$, and well agreed with the classical capillary wave theory for a rough type interface; b) the impact from the short-range structural force on this wave vector range is negligible. We perform weighted least-squares fits of $\Gamma_{sp,ps}$ within this $q$ range to estimate $\tilde{\gamma}$ with values of 132(4)$\times 10^{-10}$mJ/m, 119(3) $\times 10^{-10}$mJ/m, 101(5) $\times 10^{-10}$mJ/m and 57(7) $\times 10^{-10}$mJ/m for 370K, 380K, 390K and 400K, respectively. The correlated fluctuation stiffness $\Gamma_{spps}$ is determined from the cross-correlation function, Eq. (3). The magnitude of the $\Gamma_{spps}$ depicts the independence of the two bounding interfaces. As shown in Fig. 7, $\Gamma_{spps}(q)$ diverge at large $q$ for all temperatures, 400K GB system shows a clear decrease in the correlation in comparing with other three temperature systems.}

\begin{figure} [!htb]
\includegraphics[width= 3.5in]{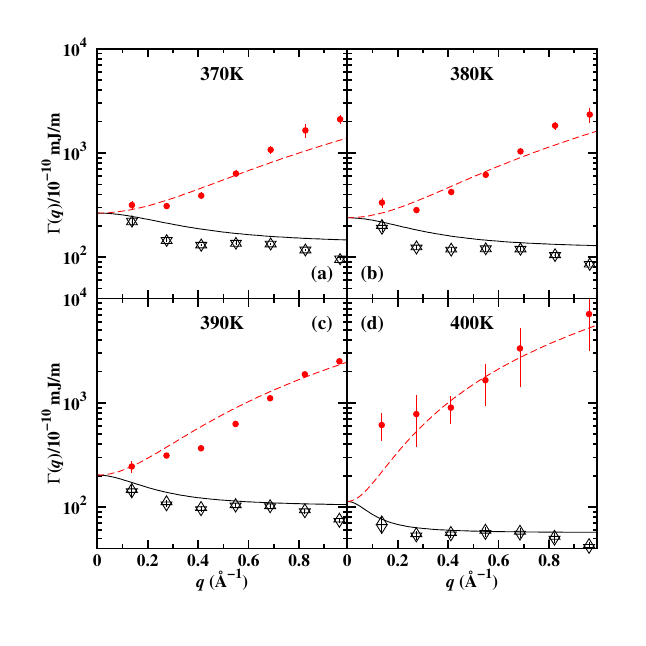}
\caption{\bl{Power spectra of equilibrium coupling fluctuations for the premelting band at the symmetric tilt GB of confined monolayer ice, misorientation angle $\theta=30^\mathrm{o}$. The plot represents wave-vector dependent stiffness, $\Gamma(q)$, in log scale for 370K (a), 380K (b), 390K (c) and 400K (d), respectively. Results for the two solid-premelt interfacial stiffness labeled with $\Gamma_{sp}(q)$ and $\Gamma_{ps}(q)$ are shown with up-pointing and down-pointing triangles, respectively. Red circles indicate the results for the coupled fluctuations of the two bounding solid-premelt interfaces, $\Gamma_{spps}(q)$. The lines are the results of weighted least-squares fits to the equation set in Eq. (2)-(3).}}
\label{fig7}
\end{figure}

\bl{In the limit of long wavelength, the fluctuation is affected by the premelting band. We find both $\Gamma_{sp,ps}$ and $\Gamma_{spps}$ asymptotically approaches to the finite value of stiffness as $q\rightarrow0$. Using the $\tilde{\gamma}(T)$ extracted from above, $\Gamma_{\alpha\beta}(q)$ fit well with Eq. (2)-(3) and $\omega=0$, indicating a rough type structure for the whole premelted GB. The fittings yield $g^{''}$ with estimated values of 15.2$\times 10^{10}$mJ/m$^3$, 10.1$\times 10^{10}$mJ/m$^3$, 4.5$\times 10^{10}$mJ/m$^3$ and 0.6$\times 10^{10}$mJ/m$^3$ for 370K, 380K, 390K and 400K, respectively. We can then examine an important length scale --- parallel correlation length ($\xi_{\|}=\sqrt{\frac{\tilde{\gamma}}{g^{''}}}$),~\cite{MacDowell14} which limits the roughness of the two bounding solid-premelt interfaces due to disjoining pressure. The estimated values of $\xi_{\|}$ are around 2.9\AA, 3.4\AA, 4.8$\AA$ and 9.9$\AA$ for 370K, 380K, 390K and 400K, respectively. These numbers are slightly smaller than the width of the premelting band, reasonably agree with the visual inspections from the simulation snapshots in Fig.2-4. Our findings suggest that premelted symmetric tilt GBs in 2D monolayer square ice are all rough from the near melting point to 40K below the melting point, in contrast to the premelting of the 3D ice surfaces in which roughening temperature is just below the melting point of bulk ice. The distinction between current premelting system and the bulk ice system could arise from different impact due to dimensionality on inhibiting the long-wavelength parallel fluctuations.}

In conclusion, using MD simulation, we predict the existence of a premelting transition at the symmetric tilt GB (misorientation angle $\theta=30^\mathrm{o}$) in monolayer ices confined in 2D hydrophobic nano-channel (channel width 6.5\AA, under a van der Waals pressure of 0.5GPa), using SPC/E model of water. The melting point (402.5K) of monolayer ice is approached from below, the interfacial region of the crystalline monolayer ice melts to form a premelting band of liquid water separating the two crystal grains. Although solid-vapor premelting in ices and GB premelting in metals and binary alloys are well established in the literature, premelting of a 2D confined ice GB has not, to our knowledge, been previously reported. Over the temperature range from 360K to 400K, the width of the premelting band is shown to depend logarithmically on the undercooling, as predicted by theoretical considerations in which the disjoining potential between the two solid-liquid (solid-premelt) interfaces is repulsive. \bl{Through the calculation of power spectra of equilibrium coupling fluctuations, we have proved the validity of a recently extended capillary wave theory for premelting fluctuation and demonstrated both the premelted GB and the two solid-premelt interfaces bounding the premelting band has the rough type structure in a broad range of undercooling temperature. Our findings are inconsistent with the fact that premelted bulk ice surface cannot be as easily roughened.~\cite{Benet16} It is believed that system dimensionality can have a significant effect on roughening behaviors of premelted ice surfaces or interfaces.} Note that, to confirm the full validity of the existence and the type of the premelting transition, comparative studies employing more realistic water models\cite{Vega11,Espinosa16} are warranted. It is highly probable that the GB premelting exists over a wide range of 2D or semi-2D confined ice phases and figures prominently in the grain coalescence, sintering, coarsening, transport behavior and many other important properties (like dielectric property\cite{Fumagalli18}) of the confined ice crystallites.

\section{Acknowledgements}
YY and HTL acknowledge Fundamental Research Funds for the Central Universities, funding from the Chinese National Science Foundation [Grant No.11504110, No.11874147] and Large Instruments Open Foundation of ECNU.
 
\newpage

\end{document}